\documentclass[%
 reprint,
nofootinbib,
bibnotes,
 amsmath,amssymb,
]{revtex4-2}
\usepackage[usenames]{color}
\usepackage {graphicx}
\usepackage{bm}

\begin{document}

\title{Two-Time relativistic Bohmian Model of Quantum Mechanics}

\author{Giuseppe Ragun{\'i}}
\email{giuseppe.raguni@um.es , giuseraguni@gmail.com}
\affiliation{{Departamento de Física, Universidad de Murcia, E-30071 Murcia, Spain.}}


\begin{abstract}
Two-Time relativistic Bohmian Model (TTBM) is a theory in which the apparently paradoxical aspects of Quantum Mechanics are the effect of the existence of an extra unobservable time dimension. The hypothesis that matter is capable of motion with respect to an additional independent time (thus resulting instantaneous with respect to usual time) is capable of restoring determinism, explaining the \emph{Zitterbewegung} without evoking virtual  antimatter. The model also predicts a relativistic correction of the uncertainty principle.

Here the model is first summarized (definition, salient properties and empiricism) and after applied to a generic spherical atomic orbit, obtaining electron oscillations in the new time dimension, $\tau$, which demonstrate the static nature of the orbitals. Something very similar happens in the case of a particle in a box, where $\tau$-oscillations cause the particle to spread out at steady states. Some astrophysical and about spin speculations follow. Finally, it is discussed how the model fits into the fundamental problem of the definition of time in Quantum Mechanics.

\medskip{}

\end{abstract}


\maketitle

\section{Introduction} In Feynman's sum of histories interpretation \cite{1,2,3} - a formulation equivalent to the standard one - all the apparent paradoxes of Quantum Mechanics converge into a single surprising aspect: the ubiquity of matter. Yet, to resolve it and restore determinism, all that's needed is to introduce an extra temporal dimension: a measure that, though shocking at first glance, is physically equivalent to the introduction of additional spatial dimensions, already contemplated by several modern theories. This is achieved in the recently introduced TTBM theory \cite{4,5}, linked to the attempts at relativistic generalization of the de Broglie-Bohm mechanics \cite{6,7,8,9,10,11}. The crucial point in this effort is
the definition of simultaneity \cite{12}. Until a couple of years ago, the most successful strategy for this purpose was to introduce models which, while respecting Lorentz invariance, must admit superluminal speeds and complex space-time structures \cite{13,14,15,16,17,18,19,20}. The approach of TTBM is more elementary: an additional time $\tau$, independent of the usual time $t$ is introduced. Motions in $\tau$ are instantaneous with
respect to \emph{t} time and, while not directly
observable, are supposed to generate the quantum uncertainties of the observables. They are caused by a Bohmian-type Quantum Potential and find the basic example
in the famous \emph{Zitterbewegung} motion \cite{21,22}, of which a more general law, free from the call of antimatter, is
achieved. Beyond the empiricism of standard Quantum Mechanics, the model foresees a
relativistic anisotropic modification of the uncertainty principle, which could be tested in
high-energy accelerators.

Here, in the following four sections, we will delve into many important aspects, both theoretical and empirical, that have only been sketched out or omitted in previous papers. In the section V, the $\tau$-movements of the atomic electrons are determined, which are able to demonstrate the origin and stability of the orbitals. Particles in box and a possible astrophysical consequence are covered in section VI, while the VII one outlines some basic points about spin in relation to the model. This is followed by section VIII, which contains the aid of TTBM in the context of the issue of defining time in Quantum Mechanics, and Section IX with final remarks.

\section{TTBM definition and properties}

Two-time physics has been considered in various contexts. An additional temporal parameter is considered, e.g., in \cite{23,24,25,26,27,28,29} and, with a different approach, in \cite{30,31,32,33,34,35} in order to more accurately explain different quantum properties or obtain supersymmetries. 

In TTBM, the new time variable, called $\tau$, is limited to causing uncertainty on a measure of $t$, otherwise being independent of it. 
If \emph{$X(t,\tau)$}  is a coordinate of the particle, we have two velocities, one \emph{classic} and one \emph{intrinsic}: 
\begin{equation}
v_{c_x}\equiv\frac{\partial X}{\partial t}\quad\quad v_{i_x}\equiv\frac{\partial X}{\partial\tau},
\end{equation}
with ${\rm d}X=v_{c_x}{\rm d}t+v_{i_x}{\rm d}\tau$. If we suppose that $v_{c_x}$ and $v_{i_x}$ depend, respectively, only
on $t$ and $\tau$\footnote{If this were not the case, classical dynamics would have long since suggested the existence of $\tau$ time.}, we get
\begin{equation}
X(t,\tau)=X_{t_0}+X_{\tau_0}+\intop_{0}^{t}v_{c_x}(\xi)\,{\rm d}\xi+\intop_{0}^{\tau}v_{i_x}(\zeta)\,{\rm d}\zeta,
\end{equation}
setting $t_0=\tau_0=0$.
Due to its total independence
on \emph{t}, $\tau$ is \emph{hidden}: the intrinsic motions are not
directly observable and are perceived as \emph{instantaneous} with
respect to time \emph{t}, measured in any reference system, for any
value of the intrinsic velocity.
Naturally, this simultaneity is totally different from the one that can exist between two $t$-events,  considered in Special Relativity. Here, the generic hidden motion occurs with respect to a new and independent temporal parameter and therefore, as long as the object does not interact with something to constitute a $t$-event, it must be considered as simultaneous in any reference system $(x,y,x,t)$. 

 Eq. (2) allows to the particle, observed in its trajectory as a function of \emph{t},
to jump instantly from one point of space to another, in principle
arbitrarily distant, through the time dimension $\tau$; and this has to be generalized in every direction of space. Actually, Quantum Mechanics has accustomed us to these surprising facts as the spreading of the electrons in their atomic orbitals and the quantum diffractions. Really, it should be like this in \emph{any motion}: according to Feynman's sum over histories interpretation, equivalent to the standard one, the physical event "the particle (or photon) moves from \emph{A} to \emph{B}" is only correctly explained by admitting that it has traveled \emph{all the possible paths from A to B at once}, also interacting with every possible virtual particle of the vacuum \cite{2,3}. The puzzling aspects of Quantum Mechanics are rooted in ubiquity and the consideration of an extra time is capable of explaining it.

The theory starts from the wavefunction
\begin{equation}
\psi(\vec{r},t,\tau)=R(\vec{r},t,\tau)\,{\rm e}^{\frac{{\rm i}}{\hbar}S(\vec{r},t,\tau)},
\end{equation}
with \emph{$R>0$} and $S$ real functions, admitting the validity of the Klein-Gordon
equation. Beginning with the free case $(V=0)$, by taking the gradient and then the divergence of eq. (3) and, separately, differentiating it twice with respect to $t$, one can obtain \cite{4}

\begin{equation}
H_{free}^{2}=m^{2}\gamma_{c}^{2}c^{4}+\hbar{}^{2}(\frac{1}{R}\frac{\partial^{2}R}{\partial t^{2}}-c^{2}\frac{\nabla{}^{2}R}{R}),
\end{equation}

\begin{equation}
\frac{\partial(R^{2}H_{free})}{\partial t}+c^{2}\vec{\nabla}\cdot(R^{2}\vec{p}_{c})=0,
\end{equation}
having assumed $\vec{\nabla}S=\vec{p}_{c}=m\gamma_{c}\vec{v}_{c}$ and
$\frac{\partial S}{\partial t}=-H_{free}$. If we set $H_{free}\equiv m\gamma_{c}c^{2}+V_{Q}$ to define the Quantum Potential $V_{Q}$, eq. (4)
attests that it is not zero
in general.

In presence of a scalar potential $V\neq0$, one has:
$H=m\gamma_{c}c^{2}+V_{Q}+V$ and eqs. (4, 5) continue to hold with $H-V$
replacing $H_{free}$, by using the generalized Klein-Gordon equation.

Eq. (5) is nothing other than the well-known continuity equation that can also be obtained by applying the Klein-Gordon equation to $\psi$ and $\psi^*$ and which fails to represent the conservation of a probability density (see, e.g. \cite{36}). However,  \emph{in TTBM the non-relativistic meaning of $R^2$ as volumetric probability density can be recovered}. For that, it is enough to \emph{expressly} request the validity of the continuity equation:

\begin{equation}
\frac{\partial R^{2}}{\partial t}+\vec{\nabla}\cdot (R^{2}\vec{v}_{c})=0,
\end{equation}
which imposes the conservation of the probability of finding the particle in a unit volume. This equation must also be extended to the relativistic case, because it is independent of the relativistic mass of the particle and could depend only on its dimensional Lorentzian contraction, which is neglected when treating the particle as a material point.

Now, 
eqs. (5) and (6)  give (see Appendix)
\begin{equation}
R^{2}\,\frac{{\rm d}\,(m\gamma_{c}c^{2})}{{\rm d}t}+\frac{\partial(R^{2}V_{Q})}{\partial t}=0.
\end{equation}

This equation does not give rise to any contradiction if $R$ (and therefore also $V_Q$)  depends on $t$ time, as we have assumed, and relates the temporal variation of the Quantum Potential to that of the relativistic mass.

Consequently, in TTBM, \emph{the Klein-Gordon equation is compatible with the Schrödingerian meaning of $R^2$ and  its conservation}, contrary to the standard model. 

Eq. (6)  simplifies when the variability of $\vec{v}_c$ from space is not considered:

\begin{equation}
\frac{\partial R^{2}}{\partial t}+\vec{v}_{c}\cdot \vec{\nabla}R^{2}=0,
\end{equation}
and holds for any function of any-order derivative of $R$ over $t$ or space.

The last equation required by the model is a wave equation for the generic $\tau$-motion:

\begin{equation}
\frac{\partial R^{2}}{\partial\tau}-v_{i_s}\frac{\partial R^{2}}{\partial s}=0,
\end{equation}
for any direction $\hat {s}$. We will find that it, together with the conservation of energy, causes an \emph{oscillatory motion} in $\tau$. Eq. (9) describes a regressive wave, moving in a mirror-like manner with respect to the $\tau$-motion of the particle and its justification is consequent to the interpretation that the model assigns to the intrinsic motions: the operation of finding the particle in a unit volume (whose probability is $R^2$) specifies a $\bar t$-event and any $\bar t$-event has uncertainty of half a period of the intrinsic oscillation of the particle. This effectively means that the particle must perform half an oscillation - the cause of uncertainty - before the measurement is achievable, ending up in antiphase with $R^2$ \cite{4}.

Since $v_{i_s}$
depends only on $\tau$, eq. (9) holds for any function of any-order
derivative of \emph{R} over \emph{t} or space. 

The motion of the particle is decomposed into a classical and a purely quantum part.
The law of classical motion is well known: if \emph{V}
is conservative, the classic total energy is constant

\begin{equation}
m\gamma_{c}c^{2}+V=const\equiv E_{c},
\end{equation}
and deriving over time one gets

\begin{equation}
\frac{{\rm d}\vec{p}_{c}}{{\rm d}t}=-\vec{\nabla}V.
\end{equation}

As for intrinsic motions, it can be demonstrated that the very hypothesis that they are the direct cause of quantum indeterminacies implies  that Special Relativity cannot be valid for them \cite{4}. Then, since the mass experiences only the relativistic increase
due to $\vec{v}_{c}$, the conservation of energy for the generic $\tau$-motion reads

\begin{equation}
V_{Q}+\frac{1}{2}m\gamma_{c}v_{i_s}^{2}\equiv E_{q_s},
\end{equation}
that is a purely quantum total energy, independent on $\tau$.
Taking the $\frac{\partial}{\partial\tau}$ of eq. (12) and using
eq. (9) with $V_{Q}$ in place of $R^{2}$, one finds

\begin{equation}
m\gamma_{c}\frac{{\rm d}{v}_{i_s}}{{\rm d}\tau}=-\frac{\partial V_Q}{\partial s}.
\end{equation}

So in TTBM, the Bohm's original motion equation \cite{8}: $\frac{{\rm d}\vec{p}}{{\rm d}t}=-\vec{\nabla}V-\vec{\nabla}V_{Q}$
does not hold; the motion is instead described by eqs. (11) and, for each spatial direction $\hat s$, (13). These corpuscular equations are complemented by the Klein-Gordon wave equation.

After these outcomes, we can identify the expression of the two-time Hamilton's action in eq. (3), that is $S=S_t+S_\tau$, where
\begin{equation}
    S_t=\int{(\frac{-mc^2}{\gamma_c}-V_Q-V){\rm d}t}; 
S_{\tau}=\int(\frac{1}{2}m\gamma_{c} v_{i_s}^{2}-V_{Q}){\rm d}\tau.
\end{equation} 

To conclude that  $\vec{\nabla}S=\vec{\nabla} S_t= \vec{p}_c$ we have to recognize that $\vec{p}_c$ is determined to less than one period
of the intrinsic oscillation of the particle. Then $\vec{\nabla} S_\tau$ is equal to the average value of $\vec{p}_{i_s}$ in this interval, that is zero. For the same reason, it turns out that  $\frac{\partial S}{\partial t}=\frac{\partial S_t}{\partial t}$. In fact, letting $\frac{\partial }{\partial t}$ enter into the integral $S_\tau$, the only term to evaluate is $\frac{\partial V_Q}{\partial t}$. Based on eqs. (8) and (13), we obtain: $\frac{\partial V_Q}{\partial t}=m\gamma_c(v_{c_x}\frac{dv_{i_x}}{d\tau}+v_{c_y}\frac{dv_{i_y}}{d\tau}+v_{c_z}\frac{dv_{i_z}}{d\tau})$ which integrated in a period of intrinsic oscillation is zero.

We will discover that, with respect to two arbitrary different directions $\hat{s}$ and $\hat{u}$, it results $v_{i_s} \neq v_{i_u}$, although only for $v_c$ comparable with $c$. It is thus recognized that $\psi$, depending on $s$ through $S_{\tau}$, is actually a $\psi_s$: the particle comes to be represented with a different wavefunction for each direction of space.
\\

The new Bohmian model can thus be so summarized: 
\begin{enumerate}
\item The spatial coordinates of the particle have to vary as a function
of two independent temporal parameters, \emph{t }and $\tau$. Motions in $\tau$ occur in all directions; they are responsible for
quantum uncertainties and not directly observable.
\item 
The particle is represented by the wavefunction (3), where $\frac{\partial S}{\partial t}=-H=-m\gamma_{c}c^{2}-V-V_Q$ and $\vec{\nabla}S=\vec{p}_{c}$. It obeys the generalized (standard replacements: ${\rm i}\hbar\frac{\partial}{\partial t}$$\:\rightarrow$$\:{\rm i}\hbar\frac{\partial}{\partial t}-V$ and $-{\rm i}\hbar\vec{\nabla}$$\:\rightarrow$$\:-{\rm i}\hbar\vec{\nabla}-\vec{P}$) Klein-Gordon equation, maintaining the Schrödingerian  interpretation for $\psi$.

\item For $t$-motion and each $\tau$-motion the continuity eq. (6) and the wave eq. (9), respectively, hold.
\item The corpuscular Bohm's law
of motion is replaced by eqs. (11) and (13) to get, respectively, \emph{t}-motion and generic $\tau$-motion; for the latter the Special Relativity
is not valid.
 \end{enumerate}

It should also be noted that, although each $\tau$-motion does not respect the Special Relativity, the maximum value of $v_{i_s}$ is still $c$: in fact it represents the quantum uncertainty on the component $s$ of $\vec {v}_c$, which cannot be greater than $c$.
\medskip{}

Apparently, the model adds a Newtonian time to the usual four-dimensional $t$-spacetime. As we shall see in the next sections, the motions in the new time variable are normally negligible for macroscopic objects.
Conversely, for corpuscles, it happens that even assuming to know the Quantum Potential with a good approximation, due to the independence of the two times and the omnidirectionality of the $\tau$-motions, the function $\vec{r}(t,\tau)$ does not represent the commonly understood trajectory, but rather a three-dimensional object of $t$-spacetime: while $t$ remains constant, we will find that the particle can $\tau$-oscillate an arbitrary number of times in any direction. However, this is an indeterminism linked to the nonexistence of a specific function $\tau(t)$ and no longer to a conceptual dualism, as in the standard interpretation. In this model, we still have a corpuscular and a wave point of view, the latter analogous to the standard one. However, there is no incompatibility between them: openly, the wave point of view is necessary for the described constitutional insufficiency of the corpuscular description.

\subsection{Some fundamental properties}
\begin{itemize}

\item[$\boldsymbol{a.}$]
From eq. (4), being $H-V=m\,\gamma_{c}\,c^{2}+V_{Q}$, we obtain for the Quantum
Potential
\begin{equation}\begin{split}\hspace{0.9cm}V_{Q}=\\&\hspace{-0.9cm}-m\gamma_{c}c^{2}+m\gamma_{c}c{}^{2}\sqrt{1+\frac{\hbar^{2}}{m^{2}\gamma_{c}^{2}c^{4}R}(\frac{\partial^{2}R}{\partial t^{2}}-c^{2}\nabla{}^{2}R)},
\end{split}\end{equation}
basically connected to the term $\frac{\partial^{2}R}{\partial t^{2}}-c^{2}\nabla{}^{2}R$, which never is zero in all space (for photons, we will find that  it vanishes just in direction of motion). Moreover, the second addend inside the root \emph{never} is small compared to 1, even in the case of non-relativistic approximation. For a free particle, for example, we will find that $V_{Q}$ oscillates with an amplitude $\frac{1}{2}m\gamma_{c}c^{2}$. This energy is really high but it only has the effect of spreading the particle in a volume of space around states that are independent of the Quantum Potential. Such states - atomic orbits and stationary solutions in a box are examples of them - can therefore be found by the standard treatment, ignoring the Quantum Potential.

\item[$\boldsymbol{b.}$] The Bohm's
non-relativistic result \cite{8}: $V_{Q}=-\frac{\hbar{}^{2}}{2m}\frac{\nabla^{2}R}{R}$
is incorrect. It could be obtained from eq. (15) by negleging $\frac{\partial^{2}R}{\partial t^{2}}$ with respect to $c^{2}\nabla{}^{2}R$ and then considering
small the second addend inside the root; but both things are wrong: the former \emph{generally} (being possible if $\vec{v}_c\vdash\vec{\nabla}R$ and in other appropriate situations), the latter \emph{always}.
Indeed, to deduce it, Bohm used the time-dependent Schrödinger
equation ${\rm i}\hbar\frac{\partial\psi}{\partial t}=-\frac{\hbar^{2}}{2m}\nabla^{2}\psi+V\psi$;
but this equation is \emph{no longer correct} if \emph{R} depends
on space and time, because it is no longer the non-relativistic limit
of the Klein Gordon equation\footnote{Despite the important differences highlighted, the heart of this model is a Quantum Potential of the type desired by Bohm, capable of recovering determinism. Therefore it seems appropriate to keep the adjective "Bohmian" for the Theory.}.

\item[$\boldsymbol{c.}$] The non-relativistic limit is obtained simply by approximating
$\gamma_{c}$. For a
free particle, we will get a \emph{Zitterbewegung}
motion even at rest, although the case $v_{c}=0$ is singular. However, it is possible to obtain the following generalized Schrödinger equation (see Appendix):
\begin{equation}\begin{split}
-\frac{\hbar{}^{2}}{2m}\frac{\nabla^{2}\psi}{\psi}=H-V-V_{Q}-\frac{\hbar{}^{2}}{2m}\frac{\nabla^{2}R}{R}+\\+{\rm i}\frac{\hbar}{2mc^{2}R^{2}}\frac{\partial(R^{2}(m\gamma_{c}c^{2}+V_{Q}))}{\partial t},
\end{split}\end{equation}
where $\gamma_c$ has been left for easier comparison with the eq. (7). It reduces to the time-independent Schrödinger  for $V_Q=0$ ($\Rightarrow\frac{\partial^{2}R}{\partial t^{2}}=\nabla{}^{2}R=0$), from eq. (7).

\end{itemize}

\section{Free particle}
 
For a free particle, $V=0$ and the classical motion eq. (11) just informs us that $\vec{p}_c$ is constant. From eq. (7) we obtain $V_Q=\frac{k}{R^2}$, with $k$ an arbitrary constant. From eq. (15) it can be seen that $V_Q$ has the relativistic mass as a factor; this motivates the choice  $k=m\gamma_c c^2 R_M^2 f$, where $R_M$ is the maximum value of $R$ and $f$ an adimensional positive arbitrary constant\footnote{By imposition that $R(x)$ is a bell curve, one can easily find that $f\leq 0$ leads to inconsistency.}.

Using eqs. (4) and (8), after two integrations and approximating for $R\rightarrow R_{M}$, it is found \cite{4}
\begin{equation}
R\simeq R_{M}(1-\frac{f(f+2)}{2\lambda^{2}}(s-s_{M})^{2}),
\end{equation}
where $s$ is a straight line forming an angle $\theta$ with $\vec{v}_c$, $s_M$ the abscissa corresponding to $R_M$ and $\lambda\equiv\frac{\hbar}{mc\gamma_{c}\gamma_{c_s}}$ (with $\gamma_{c_s}$ referred to $v_{c_s}=v_c\ {\rm{cos}}\ \theta$).
Consequently, $V_Q$ can be approximated by the potential of a harmonic oscillator

\begin{equation}
V_{Q}\simeq m\gamma_{c}c^{2}f\,(1+\frac{f(f+2)}{\lambda^{2}}(s-s_{M})^{2}).
\end{equation}
Hence, if {\sffamily{S}} and $v_{i_s}$ are the abscissa and the intrinsic velocity of the particle along  $s$, by eq. (13) we find the following sinusoidal \emph{Zitterbewegung} motion in $\tau$

\begin{equation}
{\mathsf{S}}(\tau)\simeq {\mathsf{S}}_{\tau_0}+\Delta {\mathsf{S}}\,{\rm {sin}}(\frac{c}{\lambda}\,f\sqrt{2(f+2)}\tau)
\end{equation}
\begin{equation}
v_{i_s}(\tau)\simeq \Delta {\mathsf{S}}\,\frac{c}{\lambda}\,f\sqrt{2(f+2)}\,{\rm {cos}}(\frac{c}{\lambda}\,f\sqrt{2(f+2)}\tau).
\end{equation}
By imposing $v_{i_{s_{MAX}}}=c$, we obtain $\Delta {\mathsf{S}}=\frac{\lambda}{f\sqrt{2(f+2)}}$, which generalizes the standard \emph{Zitterbewegung} if \emph{f} is of the order of unity; however, here this motion arises by self-interaction with its own wavefunction, without the need to evoke the virtual  antiparticle. According to point 1 of the theory,  the amplitude $\Delta {\mathsf{S}}$ represents the minimum positional uncertainty in $s$ of the particle: only by means of a wave packet description is it possible to take into account larger positional uncertainties and smaller momentum uncertainties.  The uncertainty principle is therefore 
\begin{equation}
\frac{\lambda}{f\sqrt{2(f+2)}}\times m\gamma_{c}c=\frac{\hbar}{f\sqrt{2(f+2)}\gamma_{c_s}}\sim\frac{\hbar}{2\gamma_{c_s}},
\end{equation}
with a unexpected directional  Lorentz factor to divide. At ultrarelativistic speeds, the quantum uncertainties in direction of motion should be negligible.

Due to eq. (12), $V_{Q}$, and so $H$, also oscillates in $\tau$ with an amplitude $\frac{1}{2}m\gamma_{c}c^{2}$; thus this value represents
the quantum uncertainty in a measure of energy for a free particle
with minimal positional uncertainty. Multiplying by
half-period of oscillation, which expresses the quantum uncertainty in a measure of $t$-time, we find the following Time-Energy Uncertainty Relation:
\begin{equation}
\frac{1}{2}m\gamma_{c}c^{2}\times\frac{\pi\lambda}{f\sqrt{2(f+2)\,}c}=\frac{\pi\hbar}{2\,f\sqrt{2(f+2)}\gamma_{c_s}}\sim\frac{\hbar}{2\gamma_{c_s}}.
\end{equation}
The Quantum Potential is never negligible, its maximum value being at least of the order of $m\gamma_{c}c^{2}$, regardless of the value of $f$. 
However, the amplitude of the $\tau$-oscillation, i.e. the quantum positional uncertainty, is inversely proportional to the rest mass. This implies that, for macroscopic bodies, it is negligible in the instants following an initial condition of minimal positional uncertainty. For a particle of one picogram, e.g., one finds it about $15$ orders of magnitude smaller than the Compton length of the electron. Even though the positional uncertainty grows spontaneously quite fast, there is another fundamental reason that hinders the spatial spreading of a macroscopic body, as we will clarify in the next section.

The purely quantum total energy results independent of the spatial direction

\begin{equation}
E_{q_s}=E_{q}=m\gamma_{c}c^{2}(f+\frac{1}{2}),
\end{equation}
and it also represents the maximum value of $V_{Q}$, corresponding to the minimum value of \emph{R}: $R_{min}=\frac{R_{M}}{\sqrt{1+\frac{1}{2f}}}$. It is therefore concluded that there is an infinite amount of energy in the spreading volume of the particle. This should not be too surprising given the introduction of the extra time parameter and we will highlight its $virtual$ character in the next section.

An important comment concerns the
\emph{spatial relativistic anisotropy of the uncertainty principle}, due to the presence of $\gamma_{c_s}$ to divide.
For photons, for example, the uncertainty principle is null in the direction of their motion; but in any other direction this does not happen, and they have oscillatory motions in $\tau$-time, which is in agreement with the observed quantum diffraction patterns, analogous to those of any other particle. In all cases, for \emph{$\hat{s}$} tending to be perpendicular
to\emph{ $\hat{v}_{c}$}, $\gamma_{c_s}\rightarrow1$ and the standard uncertainty
principle, independent of speed, is re-established. However, the case
of \emph{exact} perpendicularity is singular: for $v_{c_s}=0$, it is found that ${v}_{i_s}= const$, rather than oscillatory \cite{4}.

For the non-relativistic ($v_{c}\ll c$) and $v_{c}\rightarrow0$
cases, it is sufficient to approximate $\gamma_{c}$ in the previous
equations (respectively, with $1+\frac{v_{c}^{2}}{2c^{2}}$ and 1),
still obtaining high frequency oscillatory motions in $\tau$. 
In $v_{c}=0$ there is another singularity: in fact, it is found that ${v}_{i_s}= const$, contrary to the case $v_{c}\rightarrow0$. 

The standard \emph{Zitterbewegung} $v_{i_s}\propto c\,e^{-\frac{2\rm{i}}{\hbar}H\,t}$,
can be reobtained by imposing that $H$ is constant and interpretating the oscillatory motion in $\tau$ as occurring in $t$ \cite{4}.

These results are able to justify the physical origin of the phase to be associated with each path in Feynman's formulation of Quantum Mechanics. The approach has been specified in \cite{37}, through a particle's oscillation forward and backward in \emph{t}-time. In our model, that type of motion arises spontaneously: it is just the \emph{Zitterbewegung} in $\tau$, represented by the extra phase factor ${\rm e}^{\frac{\rm{i}}{\hbar}S_{\tau}}$.

It must be emphasized that the Theory admits that oscillatory motions similar to those found also occur in the case of an external potential $V$. The reasons are generally valid, because $V$ modifies only indirectly, and usually slightly, the Quantum Potential. Exceptions could arise just for potentials comparable to $m\gamma_c c^2$, therefore at the quark level (in which case there would be significant variations in quantum uncertainties). Later we will explicitly find the oscillatory $\tau$-motions of an electron in a circular atomic orbit.

\section{TTBM empiricism}
TTBM predicts new effects beyond standard theory. The main ones are due to the aforementioned relativistic directional differences in the intrinsic velocities. The anisotropy of the uncertainty principle has already been discussed and could be detected by an angular test in high-energy accelerators. In the case of two-slit diffraction, it should cause a weak attenuation of the interference fringes closer to the main direction of motion. Also, there is the effect due to the finiteness of the volume of space in which the Feynman integrals must be calculated: while in standard theory this volume is unbounded, in TTBM it is represented by the positional uncertainty volume swept by the particle during its motion\footnote{This redefinition of volume, together with the introduction of the new time parameter, could make the path integral a  \emph{properly mathematical} object, unlike Feynman's original formulation.}.

However, normally these effects (especially the last one) are negligible and therefore in the most common cases the experimental predictions coincide  practically with the standard ones. Under this assumption - that is, basically, neglecting the small relativistic directional differences - we will revisit all the possibilities that can arise regarding the motion of a particle in a vacuum: evolution with self-interactions, interference (e.g. triggered by wall slits) and collapse after a detection (fig. 1).

\begin{figure*} [!ht]
\centering
\includegraphics[width=\linewidth]{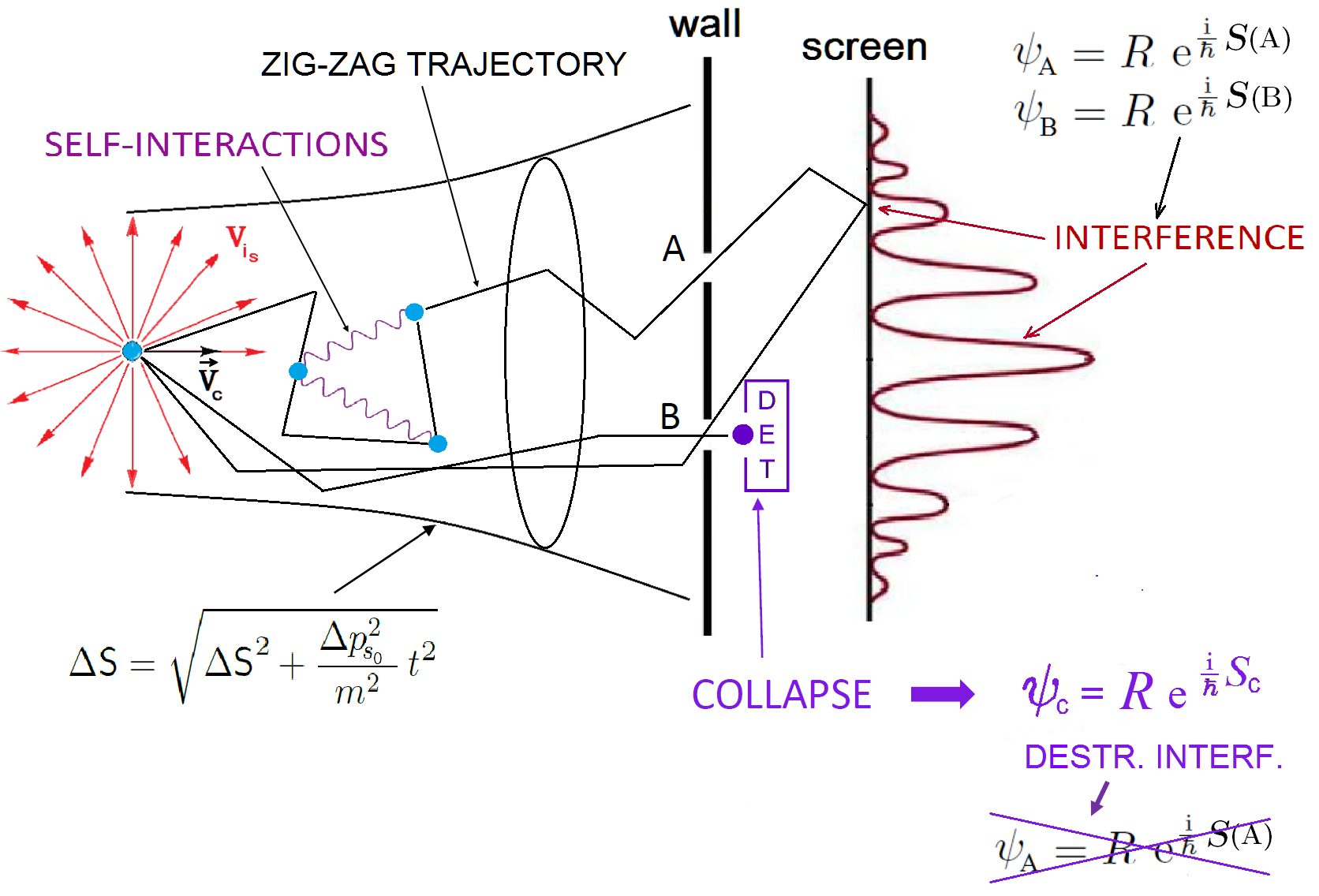}
\caption{Phenomenology of particle motion in a vacuum: self-interactions, interference (by wall slits), and observational collapse. The wavefunctions depicted are representative of packets.}
\end{figure*} 

Assuming a minimum initial positional uncertainty $\Delta {\mathsf{S}}_{0}$, as long the energy of the particle does not change during its motion, the growth of the positional uncertainty is the same as in the standard case
\begin{equation}
   \Delta {\mathsf{S}}=\sqrt{\Delta {\mathsf{S}}_{0}^{2}+\frac{\Delta p_{s_0}^{2}}{m^{2}}\,t^{2}}.
\end{equation} At a fixed instant, $\Delta {\mathsf{S}}$ represents the radius of a sphere-like figure within which the particle, due to the omnidirectional $\tau$-oscillations, is spread: relative to time  $t$, i.e. \emph{empirically}, the particle must be considered as \emph{simultaneously present in every point of this volume}.
The infinite "self-particles" interact in $\tau$-time with each other in all possible physical ways: each kind of force-mediating particles (photons, vector bosons, etc.) oscillates itself in $\tau$-time according to the same laws found. Mutual exchanges of such  particles  correspond to what in the standard point of view is described as emission and reabsorption of $virtual$ particles, obtaining excellent experimental verifications.  TTBM just specifies that \emph{the different histories are actually realized in
$\tau$-time} and therefore are instantaneous with respect to
$t$-time (i.e. with respect to an observer). The self-waves interferences happen exactly as predicted by the standard interpretation, according to the Feynman description: such interferences, constructive or destructive, are due to differences in phase, proportional to the action, caused by the different lengths of the paths.

Collapses, on the other hand, escape the standard interpretation, but can be explained in TTBM. They  occur as a consequence of an energetic change of the particle: after an inelastic impact, a subgroup of waves changes greatly its action. It predominates at the point of collision, but as it $\tau$-moves away, the subgroup thins out and, by destructively interfering with the packets having the old classic  action, it weakens, until full wave annihilation at  sufficient distance. All this happens only in $\tau$ time, so  $t$-instantaneously.

Zigzag trajectories that violate the conservation of $t$-momentum - but admitted as legitimate paths in the  Feynman interpretation - are in TTBM \emph{properly physical} (although unobservable), being obtainable by composition of $\vec p_c$ with one or more $\vec p_{i_s}$. And since the $\tau$-movements are perceived as instantaneous, this can also explain the speed of light being exceeded for certain trajectories, as admitted by the Feynman's formulation.

Of course, it is not possible to directly experience the infinite purely quantum energy instantaneously distributed in the sphere-like volume at any $t$-instant, because any measurement will cause a collapse, detecting only one particle (the only one there is!) with its finite energy.

The origin of the collapse suggests another fundamental reason that normally  prevents a macroscopic body from spread out in space through intrinsic motions: compared to a particle, it is vastly more likely that it interacts energetically with the external environment: in fact, even a small  wave phase variation  is capable of making it collapse in the way described, canceling out significant quantum uncertainties. However, if the probability of interaction is evanescent, things would be very different, as we will comment later.

Finally, we heuristically describe the case of  \emph{entangled} properties between multiple particles. According to TTBM, the state of entanglement has to be represented by a wave packet that, oscillating in $\tau$, connects all the particles however far away they may be from each other. The $\tau$-oscillation causes the result of a measurement of the property at a certain instant $\bar t$ to be predictable only probabilistically. Note that this restores full realism to the property even before any measurements, unlike de Broglie-Bohm theory itself.    A measurement of the property will cause a big phase change in a subset of wavefunctions in a certain region of space. This will result in $\tau$-interferences - in fact a kind of collapse - with the other wavefunctions of the system, capable of  influencing $t$-istantly the physical properties of any other particle.  Recall that this explicit non-locality, in hypothesis of realism, has been confirmed by experiments \cite{38,39,40}.
\section{Atomic orbitals}

Orbitals of an atom are nothing more than spread out electrons, due to their instantaneous (with respect to $t$) motion in $\tau$. In this section, we will find the oscillatory $\tau$-motions, tangential and radial, which, combined with the classical orbital $t$-motion, give rise to static orbitals (fig. 2). 

\begin{figure} [!ht]

\includegraphics[width=6.5cm, height=7.5
cm]{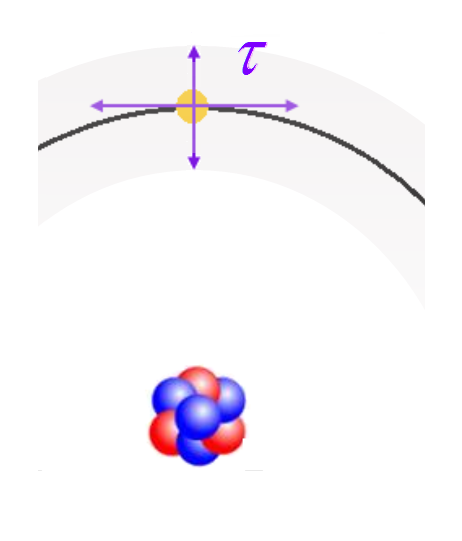}
\caption{The atomic static orbitals (shaded area) arise as a result of tangential and radial $\tau$-oscillations combined with the classical orbital motion.}
\end{figure}

The standard approach, which considers stationarity ($\frac{\partial R}{\partial t}=0$) and $V_Q=0$, is capable of determining the quantized orbits, as in a hydrogen-like atom; but only inside the new model it possible  to demonstrate, finally, the origin and static nature of the orbitals of any atom. We will limit ourselves to analyzing the case of a circular motion, so getting, in a polar reference system with center in the nucleus, just a dependence $R=R(r,\theta,t, \tau)$. 

In a small arc of an orbit of radius $\Bar{r}$, eq. (8), written for $R$, is approximated by \begin{equation}
\frac{\partial R}{\partial t}+\frac{v_c}{\Bar{r}}\dot R=0,
\end{equation}
where a dot denotes $\theta$-derivative and $\frac{{v}_c}{\Bar{r}}\equiv{\omega}_c$ is the angular velocity. Derivating further, we get
\begin{equation}
   \frac{\partial^2 R}{\partial t^2}=\ddot R\,{\omega}^2_c.
\end{equation} Due to the constancy of $v_c$, eq. (7) gives us $V_Q=\frac{{k}}{R^2}$, with ${k}$ an arbitrary constant. Eq. (4) in polar coordinates, substituting $V_Q$ and using eq. (26), becomes

\begin{equation}
\begin{split}
(\mu\gamma_{c}c^{2}+\frac{{k}}{R^2})^2-\mu^{2}\gamma_{c}^{2}c^{4}=\\\frac{\hbar{}^{2}}{R}({\omega}^2_c \ddot R-c^2(\frac{2 R'}{r}+R''+\frac{\dot R \, \rm{cos}\,\theta}{r^2 \rm{sin}\,\theta} +\frac{\ddot R}{r^2})),
\end{split}
\end{equation} where the apostrophe indicates derivative with respect to $r$ and $\mu$ is the reduced mass. At this point we have to distinguish between the cases of radial and tangential oscillation.
 
\subsection{Tangential $\tau$-oscillation}
For a small pure tangential displacement, we can neglect $R'$ and $R''$ and replace $r$, $\rm{cos}\, \theta$ and $\rm{sin}\,\theta$ with $\Bar{r}$, $1$ and $\theta$ into eq. (27), obtaining
\begin{equation}
\frac{{k}^2}{R^4}+\frac{2\mu\gamma_{c}c^{2}\, k}{R^2}=\frac{\hbar{}^{2}}{R}({\omega}^2_c-\frac{c^2}{\Bar{r}^2}) \ddot R-\frac{\hbar{}^{2}}{R}(\frac{c^2\dot R}{\Bar{r}^2\,\theta}).
\end{equation}
Substituting $\omega_c$ and simplifying, one gets the following Emden-Fowler equation

\begin{equation}
\frac{\hbar^2}{\gamma^2_c\,\Bar{r}^2}\,\theta\,\ddot R+\frac{\hbar^2}{\Bar{r}^2}\dot R+\frac{k^2}{c^2}\,\frac{\theta}{R^3}+2\mu\gamma_ck\,\frac{\theta}{R}=0.
\end{equation}
We will find the second order
approximate solution in the neighborhood of $\theta=0$, i.e.
\begin{equation}
R(\theta)\simeq R(0)+\dot{R}(0)\theta+\frac{\ddot{R}(0)}{2}\theta^{2},
\end{equation}
following the method outlined in \cite{41}.

Eq. (29) for $\theta=0$ gives $\dot R(0)=0$. After its differentiation, we get in $\theta=0$

\begin{equation}
\ddot{R}(0)=\frac{-\frac{k^2}{c^2R(0)^{3}}-\frac{2\mu\gamma_ck}{R(0)}}{\frac{\hbar^2}{\bar r^2}(1+\frac{1}{\gamma^2_c})}.
\end{equation}
Similarly to what was done for the free case, we set ${k}=\mu\gamma_c c^2 R_M^2 \Bar{f}$, where $R_M$ is the maximum value of $R$ and $\Bar{f}$ an adimensional positive arbitrary constant. Of course, not only $\Bar{f}\rightarrow f$ for $\Bar{r}\rightarrow \infty$, but more generally $\Bar{f}$ and $f$ always have to be comparable, since in any orbital the Coulomb potential is small compared to $\mu\gamma_c c^2$. Setting  $R(0)=R_M$, eq. (30) becomes
\begin{equation}
R(\theta)\simeq R_M(1-\frac{\bar f(\bar f+2)}{2\lambda^2(\gamma^2_c+1) }\bar r^2 \theta^2),
\end{equation}
where  $\lambda^{2}=\frac{\hbar{}^{2}}{\mu^{2}c^{2}\gamma_{c}^{4}}$. 
From comparison with eq. (17), taking into account that it is always $\gamma_c \simeq1$, we can conclude that again a harmonic oscillatory $\tau$-motion with characteristics very close to those obtained for the free particle is obtained.
Assuming that an electron that has just reached an empty orbit has minimum localization ($\sim\lambda$), it takes very little time to convert itself into an orbital: for $\Delta p_{s_0} \sim \mu c^2$ in eq. (24),  we find that $\Delta \mathsf{S}$ equals, e.g., the circumference of the hydrogen fundamental orbit in about $10^{-18} s$.

\subsection{Radial $\tau$-oscillation}

If in eq. (27) we require that $\theta$ remains constant and $r$ varies slightly around $\bar r$, we get the following Emden-Fowler equation: 

\begin{equation}
r\,R''(r)+2\,R'(r)+ \frac{k^2\,r}{\hbar^2c^2\,R(r)^3}+\frac{2\mu\gamma_ck\,r}{\hbar^2\,R(r)}=0.
\end{equation}
As before, we look for a second order approximate solution in the neighborhood of $r=\bar r$, i.e.
\begin{equation}
R(r)\simeq R(\bar r)+R'(\bar r)\,(r-\bar r)+\frac{R''(\bar r)}{2}(r-\bar r)^{2},
\end{equation}
but this time we directly conclude that $R'(\bar r)$ is zero, because the probability of finding the electron at $r=\bar r$ is maximum. Specifying eq. (33) in $r=\bar r$, we so get
\begin{equation}
R''(\bar r)=- \frac{k^2}{\hbar^2c^2\,R(\bar r)^3}-\frac{2\mu\gamma_ck}{\hbar^2\,R(\bar r)}.
\end{equation}
Imposing ${k}=\mu\gamma_c c^2 R_M^2 \Bar{f}$, $R(\bar r)=R_M$ and substituting into the eq. (34), one obtains
\begin{equation}
R(r)\simeq R_M(1-\frac{\bar f(\bar f+2)}{2\lambda^2\gamma^2_c } (r-\bar r)^2),
\end{equation}
which still is very similar to eq. (17). Also in radial direction we so find a harmonic $\tau$-oscillation of
the electron, capable of originating a static (for $t$-time) thickness of the orbital. Of course, this time $\Delta r$ remains small because the electron cannot move away from the quantized orbit.

\section{Particle in box. Possible astrophysical consequence}
The previous case is conceptually analogous to that of a particle in a box, provided that it hits its walls in a perfectly elastic way. Again, the standard treatment, which considers stationarity ($\frac{\partial R}{\partial t}=0$) and $V_Q=0$, gives us the possible stationary states of the particle, quantized in energy. But according to the new model, we furthermore obtain still oscillatory $\tau$-motions that spread the particle inside the box, just like an orbital. Again, they have characteristics very similar to the $\tau$-oscillations of a free particle, because the potential merely confines the particle, while inside the box it is zero.

Generalization to multiple particles is spontaneous. However, for a gas enclosed in a container, we can conclude the spreading of its molecules only if they do not collapse due to non-zero energetic interactions: therefore, if the pressure tends to zero (or the volume to infinity) and the gas can be treated as ideal.

Is it possible for the latter to happen if the "container" is a galaxy and the "particles" are cold stars? There don't seem to be any objections in principle. As you move away from the nucleus of a galaxy, the pressure decreases more and more. Here, on the basis of the model, provided that the probability of its energy variation through collisions with other matter is practically zero, there would be no impediment for cold and inert matter, of any size, to spread out through $\tau$-motions. More and more as time grows, according to eq. (24).  This spread matter, since it does not interact with anything (in particular, neither by emitting nor absorbing photons), would be completely invisible, apart from curving spacetime. However, an estimate of how much this effect would contribute to dark matter is beyond the scope of this paper.

\section{Considerations on spin}
The idea of assigning the cause of spin to the $\tau$-motions is very spontaneous and has motivated, with a certain hazard, the very choice of the adjective "intrinsic" for the $\tau$-movements. Concretely, the explanation of spin should be rooted in the new factor  $e^{\frac{\rm{i}}{\hbar}S_{\tau}}$. Unfortunately, we have not yet been able to make significant progress on this idea. If this is right, the correct and complete description of the spin states would not belong properly to a law involving only $t$-time, such as the Klein-Gordon, Dirac, Proca, Rarita-Schwinger or Pauli-Fierz equations. The results obtained from the Dirac equation - as the most important example - although extraordinarily precise for $\rm{s}=\pm \frac{1}{2}$, should just be an approximation resulting from considering
the $\tau$-oscillations as $t$-oscillations (in fact this is how the standard \emph{Zitterbewegung} turns out). Its undoubtedly \emph{acrobatic} derivation (in Pauli's words \cite{42}) could strengthen this possibility. 
Only the Klein-Gordon equation would maintain a fundamental role, because it, consistently with his dealing only with $t$-time, totally ignores the spin, implicitly assigning it a null value.

But these considerations remain for now only a conjecture, since we do not yet know how TTBM could derive, with the necessary approximations, any of the above-mentioned standard wave equations of Quantum Mechanics.

\section{TTBM and the problem of Time in Quantum Mechanics}

In Dirac's mathematical formalization of standard Quantum Mechanics, only quantities with certain mathematical properties are considered as \emph{observable}. It turns out that the traditional time $t$ cannot satisfy them \cite{43,44} and to fit it into the theory it must be treated as a classical, Newtonian, parameter, external to the theory.
Of course, this is unsatisfactory: it is equivalent to supposing the existence of a "universal clock" which, under penalty of contradiction \cite{44}, cannot be studied within the theory itself; which thus remains incomplete.

Related to this problem is the derivation and interpretation of the Time-Energy Uncertainty Relation (TEUR):

\begin{equation}
    \Delta t \ \Delta E \gtrsim\frac{\hbar}{2},
\end{equation}
for a review and discussion of which we refer to the works \cite{45,46,47} and bibliographies contained therein.
In a fundamental paper, Aharonov and Bohm have shown both that all previous claims of TEUR are not general, and that it is indeed possible to measure $E$ with infinite precision in a finite time \cite{48}. These conclusions are unquestionable because they are a consequence of the external nature of the time variable and are not incompatible with the fact that, in certain quantum systems, various forms of TEUR may be valid; as, for example, in the quantum system considered by Mandelstamm and Tamm,  where $\Delta t$ is the lifetime of a decaying system state and $\Delta E$ the uncertainty energy \cite{49}. Indeed Busch distinguished the different types of $\Delta t$ that can be considered in eq. (37) and reiterated that: 1. An observable TEUR, with a constant positive lower bound for the product of inaccuracies, is not universally valid; 2. A full-fledged quantum mechanical theory of time measurements is still waiting to be developed  \cite{50} .

According to Altaie, Hogdson and Beige “\emph{the proposal set forth by several authors of considering an intrinsic measurement of quantum time, besides having the conventional external time {\rm [}$t${\rm ]}, is compelling}” \cite{51}.
In this direction, Page and Wootters hypothesized that this intrinsic time could be represented by an appropriate dynamic variable of the system studied  \cite{52}. But which one? In more than 40 years, no example has yet been exhibited. The solid reasons argued by Unruh and Wald about the nonexistence of such an internal variable appear to have been confirmed \cite{44}.

In the same article, these authors propose an alternative that recalls the premises of TTBM: they suggest the consideration of a "\emph{non-measurable intrinsic time} [...] that provides the essential \emph{background  structure} of Quantum Mechanics”. Despite essential differences with respect to the TTBM model (a monotonic function $\tau= \tau(t)$ is admitted, $H$ is assumed constant, a non-relativistic treatment is used) their main conclusion is in accordance with it:  "\emph{in this viewpoint, an observer has access to time {\rm [}$t${\rm ]} orderings of his observations given by the label $\tau$ whose numerical values are of no significance except for the ordering they provide}" \cite{44}. Indeed, in our model the (only) link between $\tau$ and $t$ is the fact that the $\tau$-oscillation semiperiod, $T/2$, represents the uncertainty $\Delta t$. Then, time $t$ can emerge as an ordered set of "instants" $nT/2$. By the way, not only $T$ is variable, but only a non-zero energetic interaction would be able to determine the next measurable $t$ instant.

Finally, we emphasize that the intrinsic time defined in TTBM leads without ambiguity to a directional TEUR, given by (cf. eq. (22))
\begin{equation}
    \Delta t \ \Delta E \gtrsim\frac{\hbar}{2\gamma_{c_s}},
\end{equation}
where $\gamma_{c_s}$ is the Lorentz factor relative to the classical velocity component in the direction $s$.

\section{Final remarks}
The proposed theory arises as a result of taking literally Feynman's sum over history interpretation. The postulation of a new time parameter is a much simpler expedient, from a mathematical point of view, than the criterion of the Bohmian theories mentioned in the introduction, which predict a convoluted space-time and violate the spirit of Special Relativity. The theory is able to satisfactorily explain the general empiricism of Quantum Mechanics, leading interference, collapses and self-interactions (suspiciously called  $virtual$ in the standard view) to a deterministic explanation. Furthermore, TTBM provides testable effects beyond the standard viewpoint and exciting development prospects can involve explanation of spin and part of dark matter.

Of course, the theory is in its infancy and we have pointed out some gaps that need to be filled. The most important seems to be the derivation, through approximations, of the various standard spin relativistic wave equations, \emph{in primis} the Dirac one. 
\section*{Appendix}

We prove here the eqs. (7) and (16).
\medskip{}

For eq. (7): multiplying by $m\gamma_{c}c^{2}$ eq.
(6) and adding and subtracting $R^{2}\vec{\nabla}(m\gamma_{c}c^{2})\cdot\vec{v}_{c}$
we get

\begin{equation}
m\gamma_{c}c^{2}\frac{\partial R^{2}}{\partial t}-R^{2}\vec{\nabla}(m\gamma_{c}c^{2})\cdot\vec{v}_{c}+c^{2}\vec{\nabla}\cdot(R^{2}\vec{p}_{c})=0.
\end{equation}
Substituting the last term by eq. (5) and considering that $\vec{\nabla}(m\gamma_{c}c^{2})\cdot\vec{v}_{c}=\frac{{\rm d}(m\gamma_{c}c^{2})}{{\rm d}t}-\frac{\partial(m\gamma_{c}c^{2})}{\partial t}$,
one obtains

\begin{equation}
m\gamma_{c}c^{2}\frac{\partial R^{2}}{\partial t}-R^{2}\frac{{\rm d}(m\gamma_{c}c^{2})}{{\rm d}t}+R^{2}\frac{\partial(m\gamma_{c}c^{2})}{\partial t}-\frac{\partial(R^{2}H_{free})}{\partial t}=0.
\end{equation}
Now we replace $H_{free}$ with $m\gamma_{c}c^{2}+V_{Q}$: 

\begin{equation}
\begin {split}
m\gamma_{c}c^{2}\frac{\partial R^{2}}{\partial t}+R^{2}\frac{\partial(m\gamma_{c}c^{2})}{\partial t}-\frac{\partial(R^{2}m\gamma_{c}c^{2})}{\partial t}-\\
-R^{2}\frac{{\rm d}(m\gamma_{c}c^{2})}{{\rm d}t}-\frac{\partial(R^{2}V_{Q})}{\partial t}=0,
\end{split}
\end{equation}
from which, simplifying, eq. (7) follows.
\medskip{}

For eq. (16): taking the gradient and then the divergence of $\psi$, one obtains 

\begin{equation}
\frac{\nabla^{2}\psi}{\psi}=\frac{\nabla^{2}R}{R}-\frac{p^{2}}{\hbar^{2}}+\frac{{\rm i}}{R^{2}\hbar}\vec{\nabla}\cdot(R^{2}\vec{p}_{c}).
\end{equation}
Substituting the last term by eq. (5) and multiplying by $\frac{-\hbar^{2}}{2m}$,
we get

\begin{equation}
\frac{-\hbar^{2}}{2m}\frac{\nabla^{2}\psi}{\psi}=\frac{-\hbar^{2}}{2m}\frac{\nabla^{2}R}{R}+\frac{p^{2}}{2m}+{\rm i}\frac{\hbar}{2mc^{2}R^{2}}\frac{\partial(R^{2}H_{free})}{\partial t},
\end{equation}
from which, substituting $m\gamma_{c}c^{2}+V_{Q}$ for $H_{free}$ and
replacing $\frac{p^{2}}{2m}$ with $H-V_{Q}-V$, eq. (16) follows.

\end{document}